\documentstyle[sprocl]{article}
\def\beq{\begin{equation}}
\def\eeq{\end{equation}}

\def\lsim{ {\ \lower-1.2pt\vbox{\hbox{\rlap{$<$}\lower5pt\vbox{\hbox{$\sim$}
}}}\ } }

\begin{document}
\title{ PQCD ANALYSIS OF INCLUSIVE HEAVY HADRONS DECAYS}
\author{ Hoi-Lai Yu}
\address{Institute of Physics, Academia Sinica, Taipei, Taiwan 115, 
Republic of China}
\maketitle\abstracts{We develop the perturbative QCD formalism for inclusive
 heavy hadron decays.  Transverse  degrees of freedom of partons are introduced to
facilitate the factorization  of the heavy hadron decays.}

In this talk, we shall derive the PQCD factorization formula for the
semileptonic decay $B\to X_u\ell\nu$ in a rigorous way, to demonstrate how to use techniques of PQCD for heavy hadron decays.  The factorization procedures demand
the inclusion of the transverse degrees of freedom of partons at the end point of the charge lepton spectrum where an outgoing jet is present. Hence, we
have to perform the resummation of large perturbative corrections in the transversal
configuration space using the technique developed in \cite{bb}, which
is also accurate up to next-to-leading logarithms. 

We work in the rest frame of the $B$ meson, and choose the following
light-cone components for relevant momenta,
\begin{eqnarray}
P_B=(P_B^+,P_B^-,{\bf 0_\bot})\;,\;\;\;\;
p_\ell=(p_\ell^+,0,{\bf 0_\bot})\;,\;\;\;\;
p_\nu=(p_\nu^+,p_\nu^-,{\bf p_{\nu\bot}})\;,
\label{niv}
\end{eqnarray}
with $P_B^+=P_B^-=M_B/\sqrt{2}$ and $p_\nu^2=0$.
The independent variables are identified as $p_\ell^+$,
$p_\nu^-$ and $p_\nu^+$, and their relations to $E_\ell$, $q^2$ and $q_0$
are $E_\ell=p_\ell^+/\sqrt{2}$, $q^2=2p_\ell^+ p_\nu^-$ and
$q_0=(p_\ell^++p_\nu^++p_\nu^-)/\sqrt{2}$, respectively. We define
$P_b=P_B-p$ as the $b$ quark momentum, which satisfies $P_b^2
\approx M_b^2$, $M_b$ being the $b$ quark mass. $p$ is the kicks from
the light components inside the $B$ meson, which has a large plus
component $p^+$ and small transverse components ${\bf p_\bot}$.  The $b$ quark decays into a $u$ quark with momentum $P_u=P_B-p-q$.
We have distinguished the $B$ meson momentum $P_B$ from the $b$ quark 
momentum $P_b$ here. 

It is more convenient to employ the scaling variables
\begin{equation}
\ x=\frac{2E_\ell}{M_B}\;,
\ \hskip 0.2cm
\ y=\frac{q^2}{M^2_B} \;,
\ \hskip 0.2cm
\ y_0=\frac{2q_0}{M_B}\;,
\label{a4}
\end{equation}
instead of the dimensionful ones $E_\ell$, $q^2$ and $q_0$.
Note that the scaling variables are defined in terms of the $B$ meson mass
$M_B$, since we formulate the factorization according to the  $B$ meson
kinematics. For massless leptons, it is easy to show, using the momentum 
configurations defined in eq.~(\ref{niv}), that the phase space 
is given by
\begin{equation}
\ 0\leq x \leq 1,
\hskip 0.2cm
\ 0\leq y \leq x,
\hskip 0.2cm
\ \frac{y}{x}+x\leq y_0\leq y +1\;.
\label{e6}
\end{equation}

In the end point region with $x\to 1$ ($p_\ell^+\to M_B/\sqrt{2}$) and
$y\to 0$ ($p_\nu^-\to 0$), we have $y_0\to 1$ ($p_\nu^+\to 0$) and $p\to 0$.
The $u$ quark then has a large minus component $P^-_u=(1-y/x)
M_B/\sqrt{2}$ but a very small plus component $P^+_u=(1-y_0-y/x)
M_B/\sqrt{2}$, and thus a very small invariant $P^2_u=M^2_B(1-y_0+y)$, 
which forms an on-shell jet subprocess. The $u$ quark travels a long 
distance of ${\cal O}(1/\Lambda_{\rm QCD})$ before hadronized. Besides, 
the $B$ meson is dominated by soft dynamics, which is the origin of the 
soft function stated in the Introduction. The remaining 
dominant subprocess is the hard one, which contains the weak decay vertices. 
Therefore, the important contributions are factorized into the soft $(S)$, 
jet $(J)$ and hard $(H)$ subprocesses.  

The factorization formula for the inclusive semileptonic decay 
$B\to X_u\ell\nu$ is written as
\begin{eqnarray}
\frac{1}{\Gamma^{(0)}_\ell}\frac{d^3\Gamma}{dxdydy_0}
&=&M_B^2\int^{z_{\rm max}}_{z_{\rm min}}{dz} \int d^2{\bf p_\bot}
\nonumber \\
&&\times
S(z,{\bf p_\bot},\mu)J(z,P_u^-,{\bf p_\bot},\mu)
H(z,P_{u}^-,{\bf p_\bot},\mu)\;,
\label{as}
\end{eqnarray}
with the momentum fraction $z$ defined by $z=P^+_b/P^+_B=1-p^+/P^+_B$ and
$\Gamma^{(0)}_\ell=\frac{G_F^2}{16\pi^3}|V_{ub}|^2M^5_B$. $\mu$ in 
eq.~(\ref{as}) is the renormalization and factorization scale. The triple 
differential decay rate is, of course, $\mu$ independent. Note that
in the region $y\to x \sim 1$ the outgoing $u$ quark becomes
soft and eq.~(\ref{as}) fails. We shall show that contributions from
this dangerous region are suppressed by phase space. 
The upper limit of $z$ takes the value $z_{\rm max}=1$ in our analysis.
If performing the factorization according to the $b$ quark kinematics, 
one must assume $z_{\rm max}=M_B/M_b$, which is greater than 1, in order
to fill up the kinematic window. It has been 
explained \cite{ks} that $z_{\rm max}>1$ is not allowed in perturbation
theory, and is thus of nonperturbative origin. From the kinematic
constraints in eq.~(\ref{e6}) and the on-shell condition of the 
$u$-quark jet, the lower limit of $z$ should be 
$z_{\rm min}=x $, instead of $z_{\rm min}=0 $.

The tree-level expressions for the convolution factors $J$ and $H$ are
given by
\begin{eqnarray}
\ J^{(0)}&=&\delta(P_u^2)=
\delta \left(M^2_B \left[1-y_0+y-(1-z)(1-\frac{y}{x})-\frac{2 {\bf p_\bot}
\cdot{\bf p_{\nu\bot}} }{M^2_B} -\frac{{\bf p_{\bot}}^2 }{M^2_B}\right]
\right)\;,
\nonumber\\
\ H^{(0)}&\propto&(P_b\cdot p_\nu)(p_\ell\cdot P_u)
=((P_B-p)\cdot p_\nu)(p_\ell\cdot P_u)
\nonumber \\
&\propto&(x-y)\left(y_0-x-(1-z)\frac{y}{x}+\frac{2 {\bf p_{\bot}}\cdot
{\bf p_{\nu\bot}} }{M^2_B}\right)\;.
\label{tl}
\end{eqnarray}
Equation (\ref{as}) can be regarded as an expression
at the intermediate stage in the derivation of conventional factorization
theorems. If the $\bf p_\bot$ dependence in $J$ and $H$ is negligible,
the variable $\bf p_\bot$ in $S$ can be integrated over, and eq.~(\ref{as}) 
reduces to the conventional factorization formula. However, it is obvious 
from eq.~(\ref{tl}) that at least the $\bf p_\bot$ dependence in $J$ is not
negligible, especially in the end-point region. This is the reason
we introduce the transverse degrees of freedom into our analysis.

Suppose we consider higher-order corrections to eq.~(\ref{as}) from a gluon
crossing the final state cut, and route the loop momentum $\ell$ through, 
say, the jet subprocess. Without losing generality, we approximate
$J(p^++\ell^+,P_u^-+\ell^-,{\bf
p_\bot}+ {\bf \ell_\bot})\approx J(p^+,P_u^-,{\bf p_\bot}+{\bf \ell_\bot})$
according to the kinimatic relations $\ell^+<p^+$, $\ell^-<P_u^-$ and
$\ell_\bot \approx {\bf p_\bot}$. Hence, the loop integral cannot be 
performed unless the dependence of $J$ on transverse momentum is known. 
This difficulty can be removed by Fourier transform,
\begin{equation}
J(p^+,P_u^-,{\bf p_\bot}+{\bf \ell_\bot})=\int\frac{d^2{\bf b}}{(2\pi)^2}
{\tilde J}(p^+,P_u^-,{\bf b})e^{i({\bf p_\bot}+{\bf \ell_\bot})
\cdot{\bf b}}\;,
\label{ft}
\end{equation}
where the impact parameter $b$ (Fourier conjugate variable of $p_\bot$) 
measures the transverse distance travelled by the jet.
Using eq.~(\ref{ft}), the $\ell_\bot$ dependence is decoupled from the 
jet function, and the factor $e^{i{\bf \ell_\bot}\cdot{\bf b}}$ 
is absorbed into the loop integral, which can then be performed. 
Therefore, an extra factor $e^{i\ell_\bot\cdot {\bf b}}$ is associated 
with each gluon crossing the final state cut in our formalism.

To further simplify the factorization formula, we neglect those terms 
involving ${\bf p_{\nu\bot}}$ in $J$ and $H$.
This is a good approximation
for $x\to 0$ and $x\to 1$, since for $x\to 0$ contributions from 
transverse momenta are not important, and for $x\to 1$ the magnitude
${p_{\nu\bot}}= M_B\sqrt{(y_0-x-y/x)y/x}$ vanishes.
Equation~(\ref{as}) then becomes
\begin{eqnarray}
\frac{1}{\Gamma^{(0)}_\ell}\frac{d^3\Gamma}{dxdydy_0}
&=&M_B^2\int^{1}_x{dz} \int \frac{d^2{\bf b}}{(2\pi)^2}
\nonumber \\
&&\times
{\tilde S}(z,{\bf b},\mu){\tilde J}(z,P_u^-,{\bf b},\mu)
H(z,P_{u}^-,\mu)\;.
\label{asb}
\end{eqnarray}

It can be shown that
the dominant subprocesses $J$, $S$ and $H$ contain large logarithms form
radiative corrections. In particular, $J$ gives rise to double (leading)
logarithms in the end point region. These large corrections spoil the
perturbation theory and must be organized. 
One has  to
resum these large corrections up to next-to-leading logarithms. The first
step in resummation is to map out the leading regions of radiative 
corrections.
In the collinear region with the loop momentum $\ell$ parallel to $P_u$
and in the soft region with $\ell\to 0$ we can eikonalize the heavy 
b-quark line. With the eikonal approximation, the b-quark propagator is expressed as
$1/v\cdot \ell$ to the order $1/M_B$ with $v=(1,1,{\bf 0_\bot})$. Hence,
the factorization in eq.~(\ref{asb}) is in fact valid to ${\cal O}(1/M_B)$. 
The physics involved in this approximation is that a soft gluon or a gluon 
moving parallel to $P_u$ can not explore the details of the $b$ quark, and 
its dynamics can be factorized. This is consistent with the HQEFT, where the
$b$ quark is treated as a classical relativistic particle carrying color
source. Since the large mass $M_B$ does not appear in the eikonal
propragator, the only large scale in $J$ is $P_u^-$. After eikonalization procedures, one obtain the Sudakav exponent that resumes large double logarithms arise from both collinear and soft regions, 
\begin{eqnarray}
& &S=2\int_{1/b}^{m_B}\frac{dp}{p}\int_{1/b}^{p}\frac{d\mu}{\mu}
A(\alpha_s(\mu))\;, \label{a}\\
& &A={\cal C}_F\frac{\alpha_s}{\pi}
+\left[\frac{67}{9}-\frac{\pi^2}{3}-\frac{10}{27}n_f\right]
\left(\frac{\alpha_s}{\pi}\right)^2\;,
\end{eqnarray}
with ${\cal C}_F=4/3$ the color factor.

Having summed up the double logarithms, we concentrate on the single 
logarithms in $\tilde S$, $H$ and the initial condition ${\tilde J}(b,\mu)$.
Since both the differential decay rate and the Sudakov exponent
$s(P_u^-,b)$ are RG invariant, we have the following RG equations:
\begin{eqnarray}
{\cal D}{\tilde J}(b,\mu)&=& -2\gamma_q {\tilde J}(b,\mu)\;,
\nonumber \\
{\cal D}{\tilde S}(b,\mu)&=& -\gamma_S{\tilde S}(b,\mu)\;,
\nonumber \\
{\cal D}H(P_u^-,\mu)&=& (2\gamma_q+\gamma_S)H(P_u^-,\mu)\;,
\label{ter}
\end{eqnarray}
with
\begin{equation}
\ {\cal D}=\mu\frac{\partial}{\partial\mu}+\beta(g)
\ \frac{\partial}{\partial g}\;.
\end{equation}
$\gamma_q=-\alpha_s/\pi$ is the quark anomalous dimension in axial
gauge, and $\gamma_S=-(\alpha_s/\pi)C_F$ is the anomalous dimension of 
$\tilde S$. Integrating eq.~(\ref{ter}), we obtain the evolution of all the convolution 
factors,
\begin{eqnarray}
{\tilde J}(z,P_u^-,b,\mu)&=& {\rm exp}\left[-2s(P_u^-,b)-2\int^\mu_{1/b}
\frac{d{\bar\mu}}{{\bar\mu}}\gamma_q(\alpha_s(\bar\mu))\right]
{\tilde J}(z,b,1/b)\;,
\nonumber\\
{\tilde S}(z,b,\mu)&=& {\rm exp}\left[-\int^\mu_{1/b}
\frac{d{\bar\mu}}{{\bar\mu}}\gamma_S(\alpha_s(\bar\mu))\right]f(z,b,1/b)\;,
\nonumber\\
H(z,P_u^-,\mu)&=& {\rm exp}\left[-\int^{P_u^-}_\mu
\frac{d{\bar\mu}}{{\bar\mu}}[2\gamma_q(\alpha_s(\bar\mu))
+\gamma_S(\alpha_s(\bar\mu))]\right]H(z,P_u^-,P_u^-)\;.
\label{un}
\end{eqnarray}
We shall neglect the intrinsic $b$ dependence of the distribution function 
$f$ below. If Sudakov suppression in the large-$b$ region is strong,
we may drop the evolution of $f$ and $\tilde J$ in $b$, which is proportional 
to $\alpha_s(1/b)$. Hence, we assume $f(z,b,1/b)= f(z)$,
${\tilde J}(z,b,1/b)={\tilde J}^{(0)}(z,b)$, the Fourier 
transform of the tree-level expression in eq.~(\ref{tl}), and
$H(z,P_u^-,P_u^-)=H^{(0)}(z,P_u^-)$.

Substituting eq.~(\ref{un}) into (\ref{asb}), we derive the
factorization formula for the inclusive semileptonic $B$ meson decay,
\begin{eqnarray}
\frac{1}{\Gamma^{(0)}_\ell}\frac{d^3\Gamma}{dxdydy_0}
&=&M_B^2\int^{1}_{x}
dz\int_0^{\infty}\frac{bdb}{2\pi}f(z){\tilde J}^{(0)}(z,b)
H^{(0)}(z,P_{u}^-)
\nonumber \\
&&\times
\exp[-S(P_u^-,b)]\;.
\label{as1}
\end{eqnarray}
The complete Sudakov exponent $S$ is given by
\begin{equation}
\ S(P_u^-,b)=2s(P_u^-,b)-\frac{5}{3\beta_1}{\rm ln}\frac{\hat P_u^-}
{\hat b}\;,
\label{qu}
\end{equation}
with $\hat P_u^-=\ln(P_u^-/\Lambda)$, which 
combines all the exponents in eq.~(\ref{un}) and
includes both leading and next-to-leading logarithms.
It is straightforward to observe from eq.~(\ref{qu}) that the Sudakov
form factor $e^{-S}$ falls off quickly at large $b\sim 1/
\Lambda$, where $\alpha_s(1/b) > 1$ and perturbation theory fails. 
Hence, the Sudakov form factor guarantees that  main contributions to the
factorization formula come from the small $b$, or short-distance, region,
and the perturbative treatment is indeed self-consistent.
We stress that our formalism is applicable to the entire range of the
spectrum.

Since all the double logarithmic corrections have been absorbed into the
jet subprocess, the soft function $\tilde S$ contains
only soft single logarithms. These single logarithms can be summed by solving the RG equation
${\cal D}{\tilde S}=-\gamma_S {\tilde S}$
as shown in eq.~(\ref{ter}). The solution has been given in eq.~(\ref{un}),
\begin{eqnarray}
{\tilde S}(z,b,\mu)= {\rm exp}\left[-\int^\mu_{1/b}
\frac{d{\bar\mu}}{{\bar\mu}}\gamma_S(\alpha_s(\bar\mu))\right]f(z,b)\;,
\label{de}
\end{eqnarray}
where the initial condition $f(z)$ for the evolution of $\tilde S$ must be 
determined phenomenologically. From the definition of
$\tilde S$, it is obvious that $f$ depends only on the properties of the bound
state $|B\rangle$, but not on the particular short distance subprocess. 
Therefore, $f$ is a process-independent universal function describing the
distribution of the $b$ quark inside a $B$ meson. Since the intrinsic $b$ dependence (the perturbative
$b$ dependence has been collected into the exponent $S$) 
is not known yet, we take the ansatz $f(z,b)=f(z)\exp(-\Sigma(b))$,
which leads to 
$\Sigma\to 0$ as $b\to 0$ according to the definition $f(z,b=0)\equiv
f(z)$. It is also natrual to assume $\Sigma >0$ for all $b$
from the viewpoint that the $b$ quark is bounded inside the $B$ meson. 
Hence, the intrinsic $b$ depedence provides further suppression.
The nonperturbative function $f(z)$, 
identified as the $B$-meson distribution function, can be
expressed as the matrix element of the $b$ quark fields, whose first
three moments are \cite{N,LY}
\begin{eqnarray}
& &\int_0^1f(z)dz=1\;,\nonumber \\
& &\int_0^1f(z)(1-z)dz=\frac{\bar\Lambda}{m_B}+
O(\Lambda_{\rm QCD}^2/m_B^2)\;,\nonumber \\
& &\int_0^1f(z)(1-z)^2dz=\frac{1}{m_B^2}\left({\bar\Lambda}^2-
\frac{\lambda_1}{3}\right)+O(\Lambda_{\rm QCD}^3/m_B^3)\;.
\label{mo}
\end{eqnarray}

Though the exponent $\Sigma$ is unknown, we can, however, extract its
leading behavior by means of the IR renormalon analysis. Note
that the perturbative Sudakov factor $e^{-S}$ in Eq.~(\ref{de}) becomes
unreliable as the transverse distance $b$ approaches $1/\Lambda_{\rm QCD}$.
Near this end point, $\alpha_s(1/b)$ diverges, and IR renormalon
contributions are significant. We reexpress the RG
result of the evolution of the distribution function, which is contained
in the second term of $S$, as 
\begin{equation}
W=\exp\left[4\pi{\cal C}_F\int\frac{d^{4}l}
{(2\pi)^{4}}\frac{v_\mu v_\nu}{(v\cdot l)^2}2\pi\delta(l^2)
\alpha_s(l_T^2)e^{i{\bf l}_T\cdot {\bf b}}N^{\mu\nu}\right]\;.
\label{p1}
\end{equation}
The loop integral corresponds to the correction from a real soft gluon 
attaching the two valence $b$ quarks, whose propagators have been 
replaced by the eikonal lines in the direction $v=(1,1,{\bf 0})$ \cite{L}.
The tensor $N^{\mu\nu}=g^{\mu\nu}-(n^\mu l^\nu+l^\mu n^\nu)/(n\cdot l)
+n^2l^\mu l^\nu/(n\cdot l)^2$ comes from the gluon propagator in 
axial gauge $n\cdot A=0$. We have set the argument of the running 
$\alpha_s$ to $l_T^2$, which is conjugate to the scale $b$ of the
distribution function.

Substituting the identity $\alpha_s(l_T^2)=\pi\int_0^{\infty}d\sigma 
\exp[-2\sigma \beta_1\ln(l_T/\Lambda_{\rm QCD})]$, $\beta_1=(33-2n_f)/12$,
into Eq.~(\ref{p1}), and performing the loop integral for
$n\propto (-1,1,{\bf 0})$ \cite{L}, we obtain
\begin{equation}
W=\exp\left[{\cal C}_F\int_0^{\infty}d\sigma
\left(\frac{b\Lambda_{\rm QCD}}{2}\right)^{2\sigma\beta_1}
\frac{\Gamma(-\sigma\beta_1)}{\Gamma(1+\sigma\beta_1)}\right]\;.
\end{equation}
It is easy to observe that the pole of $\Gamma(-\sigma\beta_1)$ 
at $\sigma\to 0$ gives the perturbative anomalous dimension of the 
distribution function appearing in Eq.~(\ref{a}). The extra poles at 
$\sigma\to 1/\beta_1$, $2/\beta_1$,..., then correspond to the IR 
renormalons, giving corrections of powers $b^2$, $b^4$,..., respectively. 
These renormalons
generate singularities, which must be compensated by the nonperturbative
power correctons in order to have a well-defined perturbative expansion. 
Using the ''minimal" ansatz \cite{ks}, {\it ie} picking up only the 
leading (left-most) renormalon contribution, we parametrize the exponent 
$\Sigma(b)$ by $\Sigma(b)=c'm_B^2b^2$, corresponding to the fact that the 
power corrections start at $O(b^2)$. Certainly, other parametrizations 
consistent with this fact are equally good.
We postulate the following two-parameter 
distribution function for the $B$ meson,
\begin{equation}
\ f_B(z)=N\frac{z(1-z)^2}{[(z-a)^2+\epsilon z]^2}\theta (1-z)\;,
\label{qun}
\end{equation}

The charged lepton spectrum for the decay 
$B\to X_u \ell\nu$ from the naive quark model is obtained by simply choosing 
$f(z)=\delta (1-z)$ and ignoring the transverse momentum dependence in
$J^{(0)}$ and $H^{(0)}$. A simple calculation leads to
\beq
\frac{1}{{\Gamma_{\ell}}^{(0)}}\frac{d \Gamma}{dx}
=\frac{x^2}{6}\left(3-2x\right)\;.
\label{eq53}
\eeq
 This spectrum does not
fall off at the end point of the spectrum, contradicting
the observed behavior of the inclusive semileptonic decays of $B$ mesons.
The discrepancy implies that the tree-level analysis is not appropriate,
especially in the end-point region where PQCD corrections are important.

The spectrum from the parton model without Sudakov suppression is obtained 
by adopting $H^{(0)}=(x-y)[y_0-x-(1-z)y/x]$ and
$P_u^2=M_B^2[1-y_0+y-(1-z)(1-y/x)]$. With integration over $y_0$, we derive
\beq
\frac{1}{{\Gamma_{\ell}}^{(0)}}\frac{d \Gamma}{d x}=
\int_0^{x}dy \int_x^{1} dz f(z) (x-y)(y+z-x)\;.
\label{e55a}
\eeq

At last, including Sudakov suppression and all other single logarithms,
we arrive at the charged lepton spectrum of the $B\to X_u\ell\nu$ decay
that takes into account both large perturbative and nonperturbative
corrections,
\begin{eqnarray}
\frac{1}{{\Gamma_{\ell}}^{(0)}}\frac{d \Gamma}{d x}&=&{M_B}
\int_0^{x}dy \int_0^{1/\Lambda}db \int_x^{1} dz f(z)(x-y) \xi 
\biggl[(z+y-x) J_1 (\xi M_B b)
\nonumber \\
&& 
\left.-\frac{2}{M_B b} \xi J_2(\xi M_B b)
+\xi^2 J_3(\xi M_B b)\right] e^{-S (P_u^-,b)}\;,
\label{e54}
\end{eqnarray}
with $\xi=\sqrt{(x-y)(z/x-1)}$ and $J_1, J_2, J_3$ are the Bessel functions of order 1,2 and 3 respectively.

We stress that our results do not violate the conclusion from HQEFT, if
they were interpreted in a proper way. To confirm this, we 
identify $P_b=(M_b^2/2P_B^-,P_B^-,{\bf 0})$ as the momentum carried 
by a free $b$ quark in factorization theorems for $B$ meson decays, where
the minus component $P_b^-$ has been set to $P_B^-$. That is, the free $b$
quark is not at rest inside the $B$ meson. We then
reexpress eq.~(\ref{e55a}) into a form similar to that in \cite{N}:
\beq
\frac{1}{{\Gamma_{\ell}}^{(0)}}\frac{d \Gamma}{d x}=
F(x)\theta\left(\frac{M_b^2}{M_B^2}-x\right)+F\left(\frac{M_b^2}{M_B^2}
\right)M(x)\;,
\label{ned}
\eeq
with $F(x)=x^2(3-2x)/6$ being the quark-model prediction derived from the
conventional approaches and
\begin{equation}
F\left(\frac{M_b^2}{M_B^2}\right)M(x)=
\int_0^{x}dy \int_x^{1} dz f(z) (x-y)(y+z-x)-
F(x)\theta\left(\frac{M_b^2}{M_B^2}-x\right)\;.
\label{nonp}
\end{equation}
The step function in eq.~(\ref{ned}) specifies the maximal $E_\ell$ 
in the decay of a free $b$ quark with the above momentum $P_b$. The 
function $M(x)$, representing nonperturbative corrections to the $b$ quark 
decay, coincides with the shape function $S(x)$ defined in \cite{N}.

We shall show that the contribution from $M(x)$ to the total decay rate
is indeed of ${\cal O}(1/M_B^2)$. Integrating eq.~(\ref{nonp}) over $x$,
we obtain
\begin{equation}
F\left(\frac{M_b^2}{M_B^2}\right)\int_0^1 M(x)dx=\frac{1}{12}
\int_0^1 dzz^4f(z)-\int_0^{M_b^2/M_B^2} F(x)dx\;.
\label{nonp1}
\end{equation}
An arbitrary structure function $f$, which possesses the same moment as in 
eq.~(\ref{mo}), can be expanded in terms of $\delta$-functions:
\begin{equation}
f(z)=\delta(1-z)-\frac{\bar \Lambda}{M_B}\delta'(1-z)
+{\cal O}({\bar\Lambda}^2/M_B^2)\;.
\label{ex}
\end{equation}
Inserting eq.~(\ref{ex}) into (\ref{nonp1}), we justify straightforwardly 
that the nonperturbative correction
\begin{eqnarray}
F\left(\frac{M_b}{M_B}\right)\int_0^1 M(x)dx&=&
\int_{M_b^2/M_B^2}^1 F(x)dx-\frac{1}{12}\frac{\bar\Lambda}{M_B}
\int_0^1 dz z^4\delta'(1-z)+{\cal O}({\bar\Lambda}^2/M_B^2)
\nonumber \\
&=&\frac{1}{3}\frac{\bar\Lambda}{M_B}-\frac{1}{3}\frac{\bar\Lambda}{M_B}
+{\cal O}({\bar\Lambda}^2/M_B^2)
\label{cancel}
\end{eqnarray}
vanishes at ${\cal O}(1/M_B)$ as concluded in \cite{N}.
In summary, the quark-model contribution from the window between 
$x=M_b^2/M_B^2$ and $x=1$, with a width of ${\cal O}(1/M_B)$, cancels the 
${\cal O}(1/M_B)$ correction from the structure function, such that the
nonperturbative correction is of ${\cal O}(1/M_B^2)$. Our result provides an explicit dynamical demonstration the validity of global quark-hadron duality.

\section*{Acknowledgments} 
I would like to thank the organizers for this well run, fruitful 
and stimulating workshop.
This work was supported in part by the National 
Science Council of ROC.

\renewcommand{\baselinestretch}{1.0}
\newcommand{\bi}{\bibitem}
{\small
%
 }

\newpage

\end{document}